\documentclass[%
superscriptaddress,
preprint,
 amsmath,amssymb,
 aps,
]{revtex4-1}

\usepackage{xcolor}
\usepackage{graphicx}
\usepackage{dcolumn}
\usepackage{bm}

\begin{document}

\preprint{APS/123-QED}

\title{
Deep underground laboratory measurement of $^{13}$C($\alpha$,$n$)$^{16}$O in the Gamow windows of the s- and i-processes
}


\author{B.~Gao~}
\affiliation{Joint department for nuclear physics, Lanzhou University and Institute of Modern Physics, Chinese Academy of Sciences, Lanzhou 730000, China}
\affiliation{CAS Key Laboratory of High Precision Nuclear Spectroscopy,
Institute of Modern Physics, Chinese Academy of Sciences, Lanzhou 73000,
People's Republic of China}
\affiliation{School of Nuclear Science and Technology, University of Chinese
Academy of Sciences, Beijing 100049, People's Republic of China}
\author{T.Y.~Jiao}
\affiliation{CAS Key Laboratory of High Precision Nuclear Spectroscopy,
Institute of Modern Physics, Chinese Academy of Sciences, Lanzhou 73000,
People's Republic of China}
\affiliation{School of Nuclear Science and Technology, University of Chinese
Academy of Sciences, Beijing 100049, People's Republic of China}
\author{Y.T.~Li}
\affiliation{CAS Key Laboratory of High Precision Nuclear Spectroscopy,
Institute of Modern Physics, Chinese Academy of Sciences, Lanzhou 73000,
People's Republic of China}
\affiliation{School of Nuclear Science and Technology, University of Chinese
Academy of Sciences, Beijing 100049, People's Republic of China}
\author{H.~Chen}
\affiliation{CAS Key Laboratory of High Precision Nuclear Spectroscopy,
Institute of Modern Physics, Chinese Academy of Sciences, Lanzhou 73000,
People's Republic of China}
\affiliation{School of Nuclear Science and Technology, University of Chinese
Academy of Sciences, Beijing 100049, People's Republic of China}
\author{W.P.~Lin}
\affiliation{Key Laboratory of Radiation Physics and Technology of the
Ministry of Education, Institute of Nuclear Science and Technology, Sichuan
University, Chengdu 610064, China}
\author{Z.~An}
\affiliation{Key Laboratory of Radiation Physics and Technology of the
Ministry of Education, Institute of Nuclear Science and Technology, Sichuan
University, Chengdu 610064, China}
\author{L.H.~Ru}
\affiliation{CAS Key Laboratory of High Precision Nuclear Spectroscopy,
Institute of Modern Physics, Chinese Academy of Sciences, Lanzhou 73000,
People's Republic of China}
\author{Z.C.~Zhang}
\affiliation{CAS Key Laboratory of High Precision Nuclear Spectroscopy,
Institute of Modern Physics, Chinese Academy of Sciences, Lanzhou 73000,
People's Republic of China}
\affiliation{School of Nuclear Science and Technology, University of Chinese
Academy of Sciences, Beijing 100049, People's Republic of China}
\author{X.D.~Tang}
\email[]{xtang@impcas.ac.cn}
\affiliation{Joint department for nuclear physics, Lanzhou University and Institute of Modern Physics, Chinese Academy of Sciences, Lanzhou 730000, China}
\affiliation{CAS Key Laboratory of High Precision Nuclear Spectroscopy,
Institute of Modern Physics, Chinese Academy of Sciences, Lanzhou 73000,
People's Republic of China}
\affiliation{School of Nuclear Science and Technology, University of Chinese
Academy of Sciences, Beijing 100049, People's Republic of China}
\author{X.Y.~Wang}
\affiliation{CAS Key Laboratory of High Precision Nuclear Spectroscopy,
Institute of Modern Physics, Chinese Academy of Sciences, Lanzhou 73000,
People's Republic of China}
\affiliation{School of Nuclear Science and Technology, University of Chinese
Academy of Sciences, Beijing 100049, People's Republic of China}
\author{N.T.~Zhang}
\affiliation{CAS Key Laboratory of High Precision Nuclear Spectroscopy,
Institute of Modern Physics, Chinese Academy of Sciences, Lanzhou 73000,
People's Republic of China}
\affiliation{School of Nuclear Science and Technology, University of Chinese
Academy of Sciences, Beijing 100049, People's Republic of China}
\author{X.~Fang}
\affiliation{Sino-French Institute of Nuclear Engineering and Technology,
Sun Yat-sen University, Zhuhai, Guangdong 519082, People's Republic of China}
\author{\textcolor{black}{D.H.~Xie}}
\affiliation{Key Laboratory of Radiation Physics and Technology of the
Ministry of Education, Institute of Nuclear Science and Technology, Sichuan
University, Chengdu 610064, China}
\author{\textcolor{black}{Y.H.~Fan}}
\affiliation{CAS Key Laboratory of High Precision Nuclear Spectroscopy,
Institute of Modern Physics, Chinese Academy of Sciences, Lanzhou 73000,
People's Republic of China}
\affiliation{School of Nuclear Science and Technology, Lanzhou University, Lanzhou 730000, China}
\author{\textcolor{black}{L.~Ma}}
\affiliation{Key Laboratory of Radiation Physics and Technology of the
Ministry of Education, Institute of Nuclear Science and Technology, Sichuan
University, Chengdu 610064, China}
\author{\textcolor{black}{X.~Zhang}}
\affiliation{Key Laboratory of Radiation Physics and Technology of the
Ministry of Education, Institute of Nuclear Science and Technology, Sichuan
University, Chengdu 610064, China}
\author{\textcolor{black}{F.~Bai}}
\affiliation{Key Laboratory of Radiation Physics and Technology of the
Ministry of Education, Institute of Nuclear Science and Technology, Sichuan
University, Chengdu 610064, China}
\author{\textcolor{black}{P.~Wang}}
\affiliation{Key Laboratory of Radiation Physics and Technology of the
Ministry of Education, Institute of Nuclear Science and Technology, Sichuan
University, Chengdu 610064, China}
\author{\textcolor{black}{Y.X.~Fan}}
\affiliation{Key Laboratory of Radiation Physics and Technology of the
Ministry of Education, Institute of Nuclear Science and Technology, Sichuan
University, Chengdu 610064, China}
\author{\textcolor{black}{G.~Liu}}
\affiliation{Key Laboratory of Radiation Physics and Technology of the
Ministry of Education, Institute of Nuclear Science and Technology, Sichuan
University, Chengdu 610064, China}
\author{H.X.~Huang}
\address{China Institute of Atomic Energy, Beijing 102413, People's Republic
of China}
\author{Q.~Wu}
\affiliation{Institute of Modern Physics, Chinese Academy of Sciences,
Lanzhou 730000, China}
\author{Y.B.~Zhu}
\affiliation{Institute of Modern Physics, Chinese Academy of Sciences,
Lanzhou 730000, China}
\author{J.L.~Chai}
\affiliation{Institute of Modern Physics, Chinese Academy of Sciences,
Lanzhou 730000, China}
\author{J.~Q.~Li}
\affiliation{Institute of Modern Physics, Chinese Academy of Sciences,
Lanzhou 730000, China}
\author{L.~T.~Sun}
\affiliation{Institute of Modern Physics, Chinese Academy of Sciences,
Lanzhou 730000, China}
\author{S.~Wang}
\affiliation{Shandong Provincial Key Laboratory of Optical Astronomy and
Solar-Terrestrial Environment, Institute of Space Sciences, Shandong
University, Weihai 264209, China}
\author{J.W.~Cai}
\affiliation{CAS Key Laboratory of High Precision Nuclear Spectroscopy,
Institute of Modern Physics, Chinese Academy of Sciences, Lanzhou 73000,
People's Republic of China}
\affiliation{School of Nuclear Science and Technology, University of Chinese
Academy of Sciences, Beijing 100049, People's Republic of China}
\author{Y.Z.~Li}
\affiliation{CAS Key Laboratory of High Precision Nuclear Spectroscopy,
Institute of Modern Physics, Chinese Academy of Sciences, Lanzhou 73000,
People's Republic of China}
\affiliation{School of Nuclear Science and Technology, University of Chinese
Academy of Sciences, Beijing 100049, People's Republic of China}
\author{J.~Su~}
\affiliation{China Institute of Atomic Energy, Beijing 102413, People's
Republic of China}
\affiliation{Key Laboratory of Beam Technology and Material Modification of
Ministry of Education, College of Nuclear Science and Technology, Beijing
Normal University, Beijing 100875, China}
\author{H.~Zhang~}
\affiliation{China Institute of Atomic Energy, Beijing 102413, People's
Republic of China}
\author{Z.~H.~Li~}
\affiliation{School of Nuclear Science and Technology, University of Chinese
Academy of Sciences, Beijing 100049, People's Republic of China}
\affiliation{China Institute of Atomic Energy, Beijing 102413, People's
Republic of China}
\author{Y.~J.~Li~}
\affiliation{China Institute of Atomic Energy, Beijing 102413, People's
Republic of China}
\author{E.~T.~Li~}
\affiliation{College of Physics and Optoelectronic Engineering, Shenzhen
University, Shenzhen 518060, China}
\author{C.~Chen~}
\affiliation{China Institute of Atomic Energy, Beijing 102413, People's
Republic of China}
\author{Y.~P.~Shen~}
\affiliation{China Institute of Atomic Energy, Beijing 102413, People's
Republic of China}
\author{G.~Lian~}
\affiliation{China Institute of Atomic Energy, Beijing 102413, People's
Republic of China}
\author{B.~Guo~}
\affiliation{China Institute of Atomic Energy, Beijing 102413, People's
Republic of China}
\author{X.~Y.~Li~}
\affiliation{Key Laboratory of Beam Technology and Material Modification of
Ministry of Education, College of Nuclear Science and Technology, Beijing
Normal University, Beijing 100875, China}
\author{L.~Y.~Zhang~}
\affiliation{Key Laboratory of Beam Technology and Material Modification of
Ministry of Education, College of Nuclear Science and Technology, Beijing
Normal University, Beijing 100875, China}
\author{J.~J.~He~}
\affiliation{Key Laboratory of Beam Technology and Material Modification of
Ministry of Education, College of Nuclear Science and Technology, Beijing
Normal University, Beijing 100875, China}
\author{Y.~D.~Sheng~}
\affiliation{Key Laboratory of Beam Technology and Material Modification of
Ministry of Education, College of Nuclear Science and Technology, Beijing
Normal University, Beijing 100875, China}
\author{Y.~J.~Chen~}
\affiliation{Key Laboratory of Beam Technology and Material Modification of
Ministry of Education, College of Nuclear Science and Technology, Beijing
Normal University, Beijing 100875, China}
\author{L.~H.~Wang~}
\affiliation{Key Laboratory of Beam Technology and Material Modification of
Ministry of Education, College of Nuclear Science and Technology, Beijing
Normal University, Beijing 100875, China}
\author{L.~Zhang~}
\affiliation{China Institute of Atomic Energy, Beijing 102413, People's
Republic of China}
\author{F.~Q.~Cao~}
\affiliation{China Institute of Atomic Energy, Beijing 102413, People's
Republic of China}
\author{W.~Nan~}
\affiliation{China Institute of Atomic Energy, Beijing 102413, People's
Republic of China}
\author{W.~K.~Nan~}
\affiliation{China Institute of Atomic Energy, Beijing 102413, People's
Republic of China}
\author{G.~X.~Li~}
\affiliation{China Institute of Atomic Energy, Beijing 102413, People's
Republic of China}
\author{N.~Song~}
\affiliation{China Institute of Atomic Energy, Beijing 102413, People's
Republic of China}
\author{B.~Q.~Cui~}
\affiliation{China Institute of Atomic Energy, Beijing 102413, People's
Republic of China}
\author{L.~H.~Chen~}
\affiliation{China Institute of Atomic Energy, Beijing 102413, People's
Republic of China}
\author{R.~G.~Ma~}
\affiliation{China Institute of Atomic Energy, Beijing 102413, People's
Republic of China}
\author{Z.~C.~Zhang}
\affiliation{College of Physics and Optoelectronic Engineering, Shenzhen
University, Shenzhen 518060, China}
\author{S.~Q.~Yan~}
\affiliation{China Institute of Atomic Energy, Beijing 102413, People's
Republic of China}
\author{J.~H.~Liao~}
\affiliation{China Institute of Atomic Energy, Beijing 102413, People's
Republic of China}
\author{Y.~B.~Wang~}
\affiliation{China Institute of Atomic Energy, Beijing 102413, People's
Republic of China}
\author{S.~Zeng~}
\affiliation{China Institute of Atomic Energy, Beijing 102413, People's
Republic of China}
\author{D.~Nan~}
\affiliation{China Institute of Atomic Energy, Beijing 102413, People's
Republic of China}
\author{Q.~W.~Fan~}
\affiliation{China Institute of Atomic Energy, Beijing 102413, People's
Republic of China}
\author{N.~C.~Qi}
\affiliation{Yalong River Hydropower Development Company, Chengdu 610051,
China}
\author{W.~L.~Sun}
\affiliation{Yalong River Hydropower Development Company, Chengdu 610051,
China}
\author{X.~Y.~Guo}
\affiliation{Yalong River Hydropower Development Company, Chengdu 610051,
China}
\author{P.~Zhang}
\affiliation{Yalong River Hydropower Development Company, Chengdu 610051,
China}
\author{Y.~H.~Chen}
\affiliation{Yalong River Hydropower Development Company, Chengdu 610051,
China}
\author{Y.~Zhou}
\affiliation{Yalong River Hydropower Development Company, Chengdu 610051,
China}
\author{J.~F.~Zhou}
\affiliation{Yalong River Hydropower Development Company, Chengdu 610051,
China}
\author{J.~R.~He}
\affiliation{Yalong River Hydropower Development Company, Chengdu 610051,
China}
\author{C.~S.~Shang}
\affiliation{Yalong River Hydropower Development Company, Chengdu 610051,
China}
\author{M.~C.~Li}
\affiliation{Yalong River Hydropower Development Company, Chengdu 610051,
China}
\author{S.~Kubono}
\affiliation{RIKEN Nishina Center, Wako, Saitama 351-0198, Japan}
\affiliation{Center for Nuclear Study, University of Tokyo, Wako, Saitama 351-0198, Japan}
\author{W.~P.~Liu~}
\email[]{wpliu@ciae.ac.cn}
\affiliation{China Institute of Atomic Energy, Beijing 102413, People's
Republic of China}


\collaboration{JUNA Collaboration}
\author{R.J.~deBoer}
\address{Department of Physics and the Joint Institute for Nuclear Astrophysics, University of Notre Dame, Notre Dame, IN 46556, USA}
\author{M.~Wiescher}
\address{Department of Physics and the Joint Institute for Nuclear Astrophysics, University of Notre Dame, Notre Dame, IN 46556, USA}
\address{Wolfson Fellow of Royal Society, School of Physics and Astronomy, University of Edinburgh, King’s Buildings,
Edinburgh EH9 3FD, United Kingdom}
\author{M.~Pignatari}
\affiliation{Konkoly Observatory, Research Centre for Astronomy and Earth Sciences (CSFK), E\"otv\"os Lor\'and Research Network (ELKH), Konkoly Thege Mikl\'{o}s \'{u}t 15-17, H-1121 Budapest, Hungary}
\affiliation{CSFK, MTA Centre of Excellence, Budapest, Konkoly Thege Mikl\'{o}s \'{u}t 15-17, H-1121, Hungary}
\affiliation{E.~A.~Milne Centre for Astrophysics, Department of Physics and Mathematics, University of Hull, HU6 7RX, United Kingdom}
\affiliation{NuGrid Collaboration, \url{http://nugridstars.org}}

\date{\today}

\begin{abstract}
The $^{13}$C($\alpha$,$n$)$^{16}$O reaction is the main neutron source for the slow-neutron-capture (s-) process in Asymptotic Giant Branch stars and for the intermediate (i-) process. 
Direct measurements at astrophysical energies in above-ground laboratories are hindered by the extremely small cross sections and vast cosmic-ray induced background. 
We performed the first consistent direct measurement in the range of $E_{\rm c.m.}=$0.24 MeV to 1.9 MeV using the accelerators at the China Jinping Underground Laboratory (CJPL) and Sichuan University. 
Our measurement covers almost the entire 
i-process Gamow window in which the large uncertainty of the previous experiments has been reduced from 60\% down to 15\%, eliminates the large systematic uncertainty in the extrapolation arising from the inconsistency of existing data sets, and provides a more reliable reaction rate for the studies of the s- and i-processes along with the first direct determination of the alpha strength for the near-threshold state.  
\end{abstract}

\maketitle


Low-mass Asymptotic Giant Branch (AGB) stars are the primary sources of elements above iron in the Galaxy via the activation of the slow-neutron-capture process (s-process)~\cite{kobayashi:20}. In these stars, the bulk of the s-process abundances is created by neutrons from the the $^{13}$C($\alpha$,$n$)$^{16}$O reaction at temperatures around $T_9$=0.1 
in the radiative $^{13}$C-pocket, located in the helium-rich layers, just below the stellar envelope~\cite{0004-637X-497-1-388,Bisterzo_2017}. 
Here $T_9$ is defined as a temperature divided by 10$^9$ Kelvin. 
The subsequent neutron capture and $\beta$-decay processes transmute lighter elements into heavier ones, with a major production efficiency between Sr and Pb depending on the initial metallicity of the star \cite{kaeppeler:11}. In some AGB simulations, it was also found that part of the $^{13}$C is still alive in the $^{13}$C-pocket at the onset of the convective thermal pulse: the remaining
$^{13}$C is mixed at the bottom of the He intershell region activating $^{13}$C($\alpha$,$n$)$^{16}$O at He burning temperatures of approximately $T_9$=0.2-0.25 \cite{Guo_2012,cristallo:18}. This scenario tends to be favored by a low $^{13}$C($\alpha$,$n$)$^{16}$O rate \cite{LUNA2021}, and the anomalous activation of the $^{13}$C($\alpha$,$n$)$^{16}$O reaction together with the $^{22}$Ne($\alpha$,$n$)$^{25}$Mg reaction at the bottom of the convective thermal pulse. This may affect the s-process isotopic pattern near the active s-process branching points \cite{cristallo:18,LUNA2021}. 

The $^{13}$C($\alpha$,$n$)$^{16}$O reaction is also the main neutron source of the intermediate process (i-process ) \cite{Cowan_1977}, which matches the puzzling abundances observed in some post AGB stars \cite{Herwig_2011}, of a subset of carbon-enhanced metal poor (CEMP) stars \cite{Hampel_2016}, of presolar grains \cite{jadhav:13} and stars in young open clusters \cite{mishenina:15}. The i-process can be activated in different stellar environments, including among other low-mass AGB stars \cite{cristallo:09,Hampel_2016}, super AGB stars \cite{jones:16} and post AGB stars \cite{Herwig_2011}, massive stars \cite{clarkson:18,Banerjee_2018} and rapidly-accreting WDs \cite{denissenkov:19}. 
In those models, a small amount of hydrogen 
is ingested into the convective helium-burning zone underneath the envelope.
Hydrogen reacts with the primary product of He-burning $^{12}$C, making $^{13}$N that will decay to $^{13}$C. The i-process is generated by the $^{13}$C($\alpha$,$n$)$^{16}$O activated at He-burning temperatures of around $T_9$=0.2 or above \cite{Herwig_2011}.  
This results in a neutron density of around 10$^{14}\sim$ 10$^{16}$ cm$^{-3}$, which is significantly higher than 
typical values of the s-process (10$^{6}\sim$ 10$^{10}$ cm$^{-3}$).
At least for one-dimensional models of metal-poor low-mass AGB models, \citet{cristallo:18} showed that after the hydrogen ingestion, the $^{13}$C($\alpha$,$n$)$^{16}$O reaction also becomes a relevant energy source in the He shell: in their calculations even a factor of two variation of the $^{13}$C($\alpha$,$n$)$^{16}$O rate changes the i-process production by orders of magnitude. Such an impact will need to be confirmed for different types of stars by multi-dimensional hydrodynamics simulations, providing guidance for how one-dimensional models should behave once hydrogen has been ingested in the hotter He-burning regions
~\cite{herwig:14,denissenkov:19,clarkson:21}.  

It is clear that the $^{13}$C($\alpha$,$n$)$^{16}$O reaction rate is a 
fundamental ingredient in the s- and i-process models, determining the neutron density and the final isotopic production. A solid understanding of the reaction cross section is needed at the associated Gamow energies of about $E_\mathrm{c.m.}$ = 0.15 to 0.3 MeV and 0.2 to 0.54 MeV, respectively. The reaction cross sections in these energy regions are strongly influenced by the $\alpha$ cluster state near the separation threshold according to the theory of Ikeda \cite{Ikeda_1972}. Descouvemont made a theoretical prediction of the level structure using a microscopic generator-coordinate method (GCM) and concluded that the theoretical S-factor below $E_\mathrm{c.m.}$ = 0.3 MeV increases rapidly with respect to the extrapolations that ignored this threshold state \cite{Descouvemont_1987}. Reliable experimental measurements down to energies below 0.3 MeV are called for to confirm the prediction. 

Considerable efforts have been devoted to push the direct measurement of the $^{13}$C($\alpha$,$n$)$^{16}$O reaction cross section ($\sigma$)  \cite{Sekharan_1967,Bair_1973,Davids_1968,Drotleff_1993,Heil_2008} down to the stellar energies where $\sigma$ becomes extremely small. 
Due to the vast cosmic background, direct measurements performed in laboratories on the Earth’s surface stopped at energies above $E_\mathrm{c.m.}$ = 0.27 MeV with a lower limit of $\sigma$ = 6(4)$\times$ $10^{-11}$ barn \cite{Drotleff_1993}, unable to 
effectively constrain the crucial threshold state and provide a reliable extrapolation down to stellar energies. 
Besides that, the extrapolation accuracy is further limited by large discrepancies among those measurements \cite{deBoer_2020, BROWN20181}. 


Recent breakthrough in the direct measurement of this reaction was reported by the LUNA collaboration \cite{LUNA2021} at $E_\mathrm{c.m.}$ = 0.23-0.30 MeV, the upper range of the s-process Gamow window.
However, they had to rely on other existing data at higher energies, ANC of the threshold state and the R-matrix analysis to determine the S-factor in the range of 0.15 to 0.5 MeV, most of which is inaccessible with the current LUNA facility. The large discrepancies between ~\citet{Harissopulos_2005} and other measurements~\cite{Drotleff_1993,Heil_2008} result in $\sim$50\% differences in their recommended upper and lower limits for the reaction rate at $T_9$=0.1-0.3, leading to significant uncertainties in the production yields of several important isotopes, such as $^{60}$Fe and $^{205}$Pb, by using their AGB model.

In this paper, we report the first consistent direct measurement of the ${}^{13}$C($\alpha$,$n$)${}^{16}$O reaction over a wider energy range of  $E_{\rm c.m.}$ = 0.24-1.9 MeV with improved precision.
Our measurement reduces the 60\% large uncertainty down to 15\% at the center of the Gamow window of the i-process, provides the first direct determination of the alpha strength for the near threshold state, and eliminates the large systematic uncertainty in the extrapolation incurred by the discrepancy of the existing experiments. A new reliable reaction rate is recommended based on our measurement.

The underground experiment was performed in the A1 hall of the China JinPing underground Laboratory (CJPL) ~\cite{doi:10.1146/annurev-nucl-102115-044842, Liu2016,JUNA2021}. 
High-intensity $^{4}$He$^+$ and $^{4}$He$^{2+}$ ions were extracted from
2.45- and 14.5-GHz electron-cyclotron-resonance (ECR) sources, 
respectively, and accelerated by a 400-kV platform called Jinping Underground Nuclear Astrophysics experimental facility (JUNA).
The highest beam energy of 800 keV was achieved by using $^{4}$He$^{2+}$ ions, allowing comparisons with previous measurements.
The acceleration voltage was calibrated using the
$^{12}$C(p,$\gamma$)$^{13}$N, $^{27}$Al(p,$\gamma$)$^{28}$Si, $^{11}$B(p,$\gamma$)$^{12}$C and $^{14}$N(p,$\gamma$)$^{15}$O
reactions \cite{Wangshuo}.  
The absolute beam energy was determined to an accuracy of 0.5 keV
 with an energy spread of less than 0.2 keV \cite{Wangshuo}.
A 90$^\circ$ dipole magnet with a mass resolution of 
250
was used together with a set of analyzing slits to eliminate the H$_2^+$/D$^+$ contamination in the $^{4}$He$^{2+}$ beam~\cite{chen18}. A clear separation of $^4$He$^{2+}$ from the other impurities was observed at the slit position and the count ratio of the inner and outer rings of the $^{3}$He detector array indicated that no neutrons came from the deuterium impurity.


To avoid the source of systematic uncertainty incurred by target deterioration in traditional thin target experiments,  
we used 2-mm thick $^{13}$C enriched targets with a purity of 97\%. The target was installed 
on a water-cooled copper backing. On-target beam intensity of 
up to 2.5 particle mA, the highest $\alpha$ beam intensity among the deep underground laboratories, was achieved. 
The thick targets turned out to be very stable and only two targets 
were used for the whole experiment. 
A cold trap 
was installed to reduce the natural carbon buildup on the targets. 
For the $^4$He$^{+}$ runs, the analyzing slits used in the $^4$He$^{2+}$ runs were removed to allow maximum transmission efficiency and achieve higher beam intensities.

Neutrons from the $^{13}$C($\alpha$, $n$)$^{16}$O reaction 
were detected by an array consisting of 24 $^{3}$He-filled proportional counters.
By placing 35-cm thick 7\% borated polyethylene blocks and 1-mm thick cadmium sheets around the detection array, a background of 4.7(2) events/hour was achieved, compared to 1238(11) events/hour measured on the Earth's surface. The detection efficiency of the array was determined to 
be 26\% for 2.5-MeV neutrons using the $^{51}$V(p,n)$^{51}$Cr reaction together with Geant4 simulations \cite{he3_array}. 

The thick target yield ${\rm Y(E)}$ of the $^{13}$C($\alpha$,$n$)$^{16}$O reaction was
measured with beam energies of 0.3 $<$ $E_\alpha$ $<$ 0.785 MeV.
The beam-induced neutron background (BINB) was estimated by measuring the 
${\rm Y(E)}$ at $E_\alpha$ = 0.25 MeV 
where the cross section of the $^{13}$C($\alpha$,$n$)$^{16}$O
reaction is negligibly small and all neutrons detected above the environmental background level should be attributed to the BINB.
BINB was determined to be 0.05(8) events/Coulomb, consistent with zero. 

The cross sections  and the corresponding effective energies were extracted by differentiating the thick target yield~\cite{Spillane_2007,Notani_2012}. We repeatedly checked the neutron yields at 17 energy points and the reproducibility was found to be 8\% and 3\% for the $^4$He$^+$ and $^4$He$^{2+}$ data sets, respectively. This random error likely originates from the beam tuning and the potential carbon build up. 
Therefore, we added this error quadratically together with the statistical error. 

Another thick target measurement was performed in the range of $E_{\rm c.m.}$ = 0.75 MeV to 1.9 MeV using the $^4$He$^+$ beam from the 3 MV Tandetron at Sichuan University~\cite{HAN201868} to resolve the discrepancies among the S-factors at higher energies in the previous works~\cite{Harissopulos_2005,Bair_1973,Drotleff_1993}. The same detection setup was used to minimize extra systematic uncertainties. 
The beam energy was calibrated using the $^7$Li(p,n)$^7$Be reaction, and confirmed by the narrow resonances of the ${}^{13}$C($\alpha$,$n$)$^{16}$O at $E_\alpha$=1055.6, 1334.7 and 1338.8 keV. A thin target measurement was also carried out in the range of $E_\mathrm{c.m.}$=1.6 to 1.9 MeV using a 3.2$\mu$g/cm$^2$-thick $^{13}$C target. The thin-target data is normalized to the thick-target data. The reproducibility of the thick-target and thin-target measurements are estimated to be 3\% and 2\%, respectively. This uncertainty is included with the statistical error as discussed above. 

By adopting a compiled angular distribution \cite{brune, PhysRev.107.1065} in the Geant4 simulation, our efficiency has been corrected for angular distribution effects. These effects were found to change the efficiency by $\pm$2\% at $E_\mathrm{c.m.} < $0.6 MeV. However, the efficiency at $E_\mathrm{c.m.}\sim$0.9 MeV deviates from the nominal efficiency with an isotropic distribution by $\sim$5\%, which is larger than the statistical uncertainties in the previous measurements~\cite{Harissopulos_2005,Bair_1973}. This deviation becomes even larger at the narrow resonances~\cite{he3_array}. Such an effect was overlooked in these previous works.

Systematic uncertainties of our measurements at CJPL and SCU are estimated to be 11\%, which includes contributions from the beam current integration(5\%), detection efficiency (7\%), angular distribution (2\% for JUNA underground measurement and 4\% for SCU experiment), and stopping power (6\%)~\cite{srim_2010}. 

\begin{figure} [htp]
    \centering
    \includegraphics[width=0.5\textwidth]{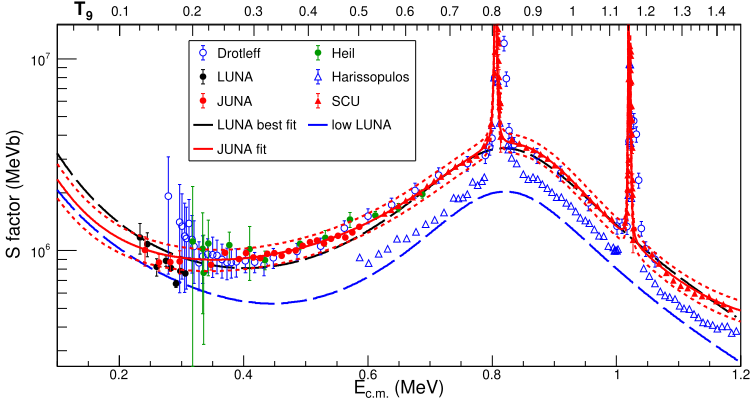} 
    \caption{The S-factor of the $^{13}$C($\alpha$,$n$)$^{16}$O reaction. The uncertainties from the fit to the JUNA+SCU data are indicated by dotted lines. The best fit and lower limit recommended by LUNA~\cite{LUNA2021} are shown as black and blue dashed lines, respectively. The S-factors have been corrected with the screening potential $U_e$ = 0.78 keV. The temperatures in $T_9$ on the top correspond to the center energy of the Gamow window on the bottom.}\label{fig_sfactor}
\end{figure}

Our S-factor is converted into the bare S-factors after correcting for the screen effect using our fitted screening potential of $U_e$ = 0.78 keV together with the previous measurements \cite{Drotleff_1993,Heil_2008,Harissopulos_2005,LUNA2021, Kellogg_1989}. The results are presented in Fig. \ref{fig_sfactor} . It can be seen that our underground data cover the energy range from $E_{\rm c.m.}$ = 0.24 to 0.59 MeV, greatly overlapping with the astrophysical region of $E_{\rm c.m.}$ = 0.15 to 0.5 MeV with a statistical uncertainty better than 15\%. 

With the unique energy range and ultra low neutron background in the deep underground lab, we are able to precisely measure the S-factor in the range of astrophysical interest for i-process nucleosynthesis. The center of the Gamow window for the i-process is located at 0.35 MeV, beyond the accessible energy range of LUNA. The two  extrapolation scenarios of LUNA using  either the normalization of Heil, or Drotleff~\cite{Drotleff_1993,Heil_2008,Bair_1973} or that of ~\citet{Harissopulos_2005}
resulted in their so-called best fit and ``low LUNA" fit, respectively. To be on the safe side, they defined the ``low-LUNA" fit by taking the 95\% confidence level of the lower limit of the fit with the original Harissopulos data. These two fits differ from each other by a factor of 2 at 0.35 MeV. Such a large systematic uncertainty in their extrapolation is eliminated by our consistent measurement, which rules out the lower normalization of~\citet{Harissopulos_2005}. ~\citet{Drotleff_1993} was the best measurement before ours at the energy around 0.35 MeV. While our data above 0.4 MeV is in good agreement with that of Drotleff, our data around 0.27 MeV are about 50\% lower and disagree with the upturning trend in this data set. The nearly 60\% uncertainty in Ref.~\cite{Drotleff_1993} within the Gamow window has been reduced to 15\%.  



The S-factor at $E_{\rm c.m.}<$0.24 MeV was obtained using an $R$-matrix analysis~\cite{RevModPhys.30.257} in the range of $E_{\rm c.m.}$=0.24 to 1.9 MeV using the code \texttt{AZURE2}~\cite{azure, azure2}. 
In our analysis, we only included our measurements of the $^{13}$C($\alpha$,$n$)$^{16}$O cross section, to eliminate the systematic uncertainty of the inconsistent data sets,  and the $^{16}$O+$n$ total cross section \cite{Fowler_1973}. 

Our best fit is shown together with its estimated uncertainty in Fig.~\ref{fig_sfactor}. The screening potential($U_e$) is fitted to be 0.78$\pm$0.43 keV. It agrees with the theoretical prediction of $U_e$=0.937 keV using the adiabatic limit while ruling out the larger prediction of $U_e$=2 keV~\cite{Trippella_2017}. Our fit is about 15\% systematically higher than the LUNA measurement~\cite{LUNA2021}. The reduced-$\chi^2$ of the LUNA data is 25 by using their best fit. It drops to 1.02 with our fit after the normalization and excluding the point at $E_{\rm c.m.}$=0.29 MeV, which is 5$\sigma$ lower than our best fit. 

Although the LUNA measurement agrees with ours within the quoted errors, the inconsistency between the measurement of \citet{Harissopulos_2005} and other measurements at higher energies leads to a $\sim$50\% difference between the upper and the lower limits of the reaction rate recommended by LUNA at $T_9$=0.1-0.3.  
This demonstrates a key limitation of the LUNA measurement, that its limited energy range did not allow for a direct comparison with higher energy data. Using our consistent measurement over a board energy range, the uncertainty of our fit are reliably constrained to the level of$<$16\% at the Gamow windows of s- and i-processes. 



\begin{figure}
    \includegraphics[width=0.45\textwidth]{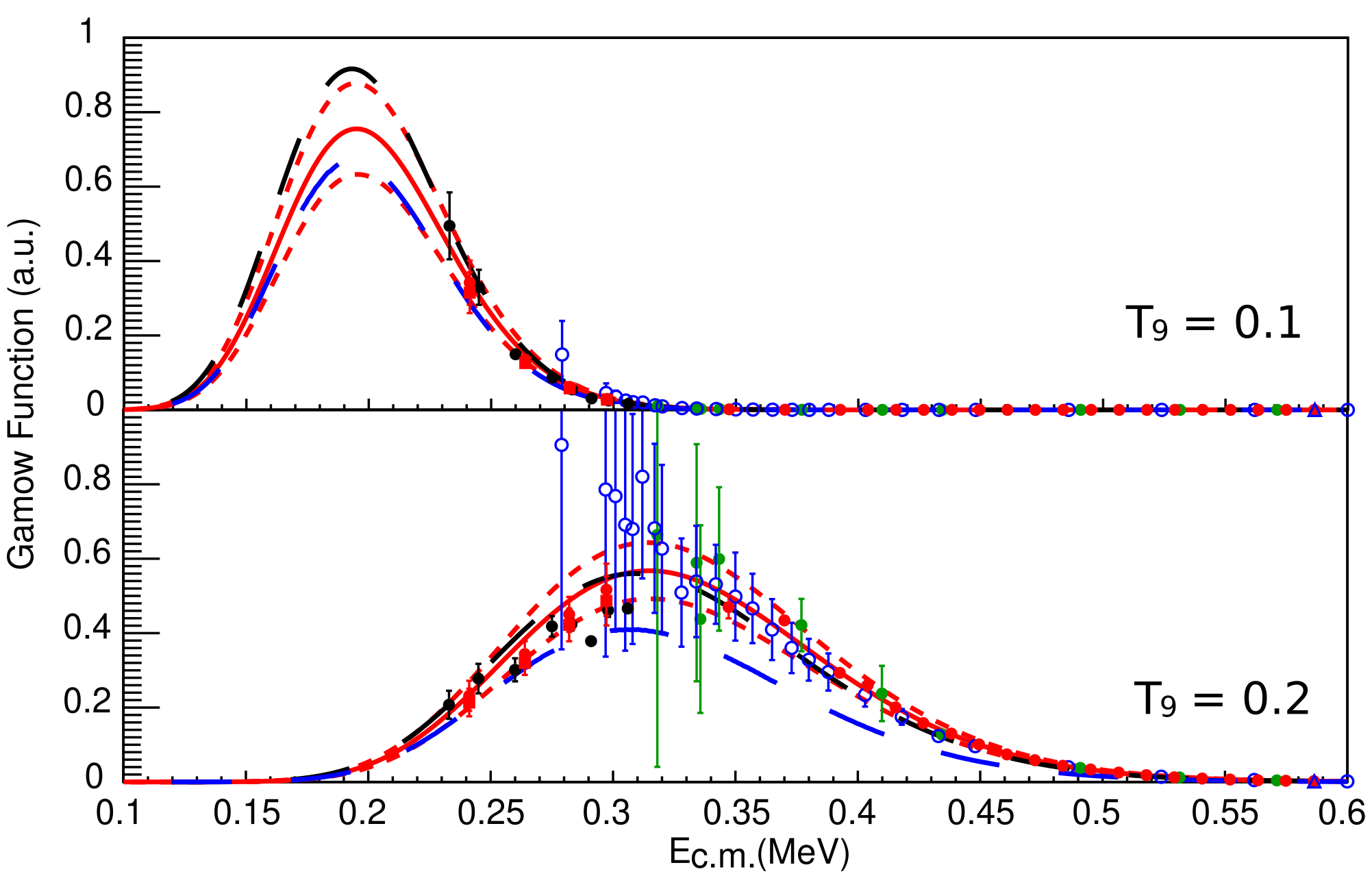} 
    \caption{The Gamow function of $^{13}$C($\alpha$,$n$)$^{16}$O at $T_9$ = 0.1 and 0.2. Color coding is identical to Fig.\ref{fig_sfactor} \label{fig_gamow}}
\end{figure}

The extrapolated S-factor towards lower energy is dominated by the $\alpha$ reduced width $\gamma_{\alpha}$ or the Coulomb renormalized asymptotic normalization coefficient ($\tilde{C}^2$) of the 1/2$^+$ threshold state. 
The R-matrix analysis performed in previous works involved fixing the ANC of the threshold state to values obtained from indirect measurements. However, the uncertainties in these ANCs often suffer from difficulties to quantify systematic uncertainties from the models used to obtain them.
The lower and higher limits of the measured $\tilde{C}^2$ differ from each other by a factor of $\sim$5~\cite{deBoer_2020}. These systematic uncertainties have been eliminated in our fit by treating the $\Gamma_{\alpha}$ of this state as a free parameter. The reduced widths $\gamma_{\alpha}$ obtained from our best $R$-matrix analysis is  -0.14(2) MeV$^{1/2}$ with a channel radius of 6.684 fm and $E_x$=6.3772 MeV, corresponding to an ANC of $\tilde{C}^2$=2.1(5) fm$^{-1}$ with $E_x$=6.356 MeV~\cite{akram2017,brune20}. Our value is slightly lower than the indirect measurements of 3.6(7) fm$^{-1}$ \cite{Avila_2015} and agree with 2.7(8) fm$^{-1}$ \cite{Guo_2012,Shen_2019} and 4.5(2.2)\cite{pellegriti08}. For the first time, we not only validate the $\alpha$ width of the threshold state obtained with the indirect method using the direct measurement, but also determine the interference pattern in the $R$-matrix analysis. As LUNA used the higher $\tilde{C}^2$ from ~\citet{Avila_2015} to constrain their extrapolation towards lower energies, our best fit is 23\% lower than their best fit at $E_{\rm c.m.}$=0.19 MeV, the center of the Gamow window for $T_9$=0.1 (see Fig.\ref{fig_gamow}). At the same energy, with the combination of a larger reduced width~\cite{Avila_2015} and the cross section of ~\citet{Harissopulos_2005}, the ``low LUNA" fit is 11\% lower than our best fit.  

The $^{13}$C($\alpha$,$n$)$^{16}$O reaction rate is calculated by numerical integration of the standard reaction rate equation~\cite{Rolfs}: 
\begin {equation} \label{eq:reaction_rate}
\langle\sigma v\rangle= \Big(\frac{8}{\pi \mu}\Big)^{1/2} \frac{1}{{kT}^{3/2}} \int_{0}^{ \infty } \sigma(E)E\exp\Big(-\frac{E}{kT}\Big)\mathrm d{\rm E}
\end {equation}
To highlight the important stellar energy range for a typical helium burning temperature $T_9$ =\,0.1 and 0.2, the integrand of Eq.~(\ref{eq:reaction_rate}) (Gamow function) is computed and shown in Fig.~\ref{fig_gamow}.
At $T_9$ = 0.1, the temperature of the $^{13}$C pocket in the AGB model, our extrapolation is lower than the best fit of LUNA and tends to agree better with their ``low LUNA" fit. At $T_9$ = 0.2, which is of importance for both the i-process and s-process nucleosynthesis in the thermal pulse in the AGB model,
our measurement covers nearly the entire Gamow function with significantly improved uncertainties.
This is a substantial improvement compared to previous measurements as the ground-based measurements from Ref. ~\cite{Drotleff_1993,Heil_2008} covered only the upper part of the Gamow window with large uncertainties while the LUNA extrapolation suffered from the inconsistencies in the absolute cross section of the higher energy measurements.
At the center of the Gamow window of $T_9$ = 0.2, our result agrees with the best fit of LUNA within our quoted uncertainty, but rules out the ``low LUNA" fit, reflecting a significant difference in the shape of our extrapolation from that of LUNA.

\begin{figure}
    \includegraphics[width=20pc]{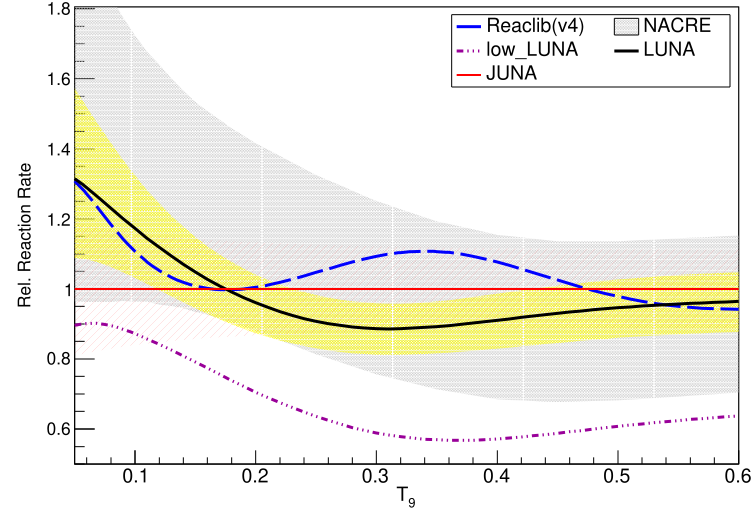}
    \caption{Selected reaction rates normalized by the rate determined in this work. The uncertainties of our new rate and the LUNA rate, based on their best fit, are indicated by the red hatched area and yellow shaded area, respectively. For comparison, we also show the rates from NACRE-II \cite{Xu_2013}, LUNA \cite{LUNA2021}, and JINA Reaclib  \cite{Guo_2012,Cyburt_2010}.}
    \label{fig:c13an_rate_comparison}
\end{figure}

The reaction rate is calculated with the JUNA fit shown in Fig.~\ref{fig_sfactor}. Comparisons of the reaction rates are shown in Fig. \ref{fig:c13an_rate_comparison}.
Our reaction rate agrees well with previous Reaclib compilations based on the ANC method  \cite{Guo_2012} and NACRE-II \cite{Xu_2013} at $T_9 \ge$ 0.1. The nearly 50\% difference between the upper limit and lower limit (``low LUNA") of LUNA and the even larger uncertainty in NACRE-II have been improved significantly. At $T_9$=0.1-0.3, the typical temperatures for s- and i-processes, we have reached an uncertainty of 13\%-16\%. 

It has been shown that the ``low-LUNA" rate 
increases the survivability of 
 $^{13}$C in the $^{13}$C-pocket of an AGB star, and that it burns at a high temperature in the subsequent thermal pulse~\cite{LUNA2021}. 
 Compared with the ``low-LUNA" rate, our recommended rate is 
slightly higher at temperatures typical of the $^{13}$C-pocket within 15$\%$, and about 30$\%$-40\% higher at the thermal pulse temperatures. Therefore, we expect our rate to produce effects similar to those discussed by \cite{LUNA2021}, although a follow up, detailed study, on AGB stellar models is needed. Concerning the impact on the i-process nucleosynthesis, future models based on the next generation of multi-dimensional hydrodynamics simulations will be more predictive thanks to the more reliable reaction rate uncertainties provided by this work.   

In summary we have performed a direct measurement of the ${}^{13}$C($\alpha$,$n$)${}^{16}$O reaction cross section over the range of $E_{\rm c.m.}$ = 0.24-1.9 MeV using the most intense $\alpha$ beam available in the deep underground laboratories with the highest precision to date. 
Our consistent measurement, covering a wide energy range, reduce the large uncertainty in the reaction rate down to 13\% to 16\% for i- and s-process nucleosynthsis.
Our reaction rate is similar to the ``low LUNA'' rate at the typical $^{13}$C pocket, favoring the release of more neutrons from the $^{13}$C($\alpha$,$n$)$^{16}$O reaction during the thermal pulse phase. Our direct measurement eliminates an important systematic uncertainty in the R-matrix extrapolation by resolving the inconsistency among the data sets at higher energies~\cite{Bair_1973,Harissopulos_2005,Kellogg_1989}. 
For the first time, we determines the ANC of the threshold state using the direct measurement, fix the interference pattern and determine the screening potential using R-matrix analysis.  

\begin{acknowledgments}
The authors would like to thank Tsinghua University and Prof. Jianping Cheng for the support of Lab infrastructure and the China National Nuclear Corporation for the finial support, CDEX and PandaX collaborations for their kind helps, L. He, Y. Wang, P. Wang and J.K. Liu for their helps in setting up the experimental terminal and Prof. Z.G. Wang for developing the instruments which was used to develop the thick $^{13}$C targets. The authors also acknowledge the assistance of Carl Brune in estimating neutron angular distribution corrections. 
X.T. thanks A. Best and D. Rapagnani for their helpful discussions throughout the project and providing their original data sets. 
This work was supported by the National Natural Science
Foundation of China under Grants No. 11490564,  the Strategic Priority Research Program of Chinese Academy of Sciences, grant No.
XDB34020000, the national key research and development
program (MOST 2016YFA0400501). S. W. was supported by the National Natural Science Foundation of China under Grants No. 11775133, 11405096. B. G. was supported by the National Natural Science Foundation of China under Grants No. 12125509. W.P. L. was supported by the National Natural Science Foundation of China under Grant No. 11805138. X. F. was supported by the National Natural Science Foundation of China under Grants No. 11875329. R.J.D. acknowledges the use of resources from the Notre Dame Center for Research Computing. R.J.D. and M.W. were supported by the National Science Foundation through Grant No. Phys-2011890, and the Joint Institute for Nuclear Astrophysics through Grant No. PHY-1430152 (JINA Center for the Evolution of the Elements). MP acknowledges support to NuGrid from STFC (through the University of Hull's Consolidated Grant ST/R000840/1) and ongoing access to {\tt viper}, the University of Hull High Performance Computing Facility. MP thanks the support by the ERC Consolidator Grant (Hungary) funding scheme (Project RADIOSTAR, G.A. n. 724560), the ChETEC COST Action (CA16117), supported by the European Cooperation in Science and Technology, the IReNA network supported by NSF AccelNet, and the European Union ChETEC-INFRA (project no. 101008324).
\end{acknowledgments}

\nocite{*}

\bibliography{c13an}

\end{document}